\documentclass[12pt]{article}
\def\be{\begin{equation}}
\def\ee{\end{equation}}
\def\bea{\begin{eqnarray}}
\def\eea{\end{eqnarray}}

\input{tcilatex}

\usepackage{graphicx}

\begin{document}

\title{An analytical solution of Monge differential equation}
\author{Rafael Torres-Cordoba \\
%EndAName
Universidad Autonoma de Cd. Juarez.\\
Av. Del Charro 450 norte, C.P. 32310\\
Cd. Juarez Chih. Mexico.}
\maketitle

\begin{abstract}
We present the exact solution to the non linear Monge differential equation $%
\lambda (x,t)\lambda _{x}(x,t)=\lambda _{t}(x,t)$. It is widely accepted
that the Monge equation is equivalent to the ODE $\stackrel{..}{X}=0$ of
free motion for particular conditions. Furthermore, the Monge Type equations
are connected with $X=F(\stackrel{.}{X,}X;t)$, which can be integrated with
quadratures \cite{leznov}. Other asymptotic solutions are discussed, see
e.g. \cite{witham}.

The solution was reached with calculations that depend upon dimensional
representation, which is given by $(x,t)$ coordinates. We present this
analytical solution to the Monge differential equation as an implicit
solution.
\end{abstract}

\section{Introduction}

The Monge equation has been cited in several textbooks for approximately 150
years so far \cite{witham}, \cite{fairlie}, and \cite{leznov}. Despite its
wide number of analytical applications (e.g. Monge-Ampere, which is in a
more general form than Monge equation itself), it has only been solved to
fulfill particular mathematical conditions. The Monge equation currently
exists in the following form:

\begin{equation}
\lambda (x,t)\frac{\partial \lambda (x,t)}{\partial x}=\frac{\partial
\lambda (x,t)}{\partial t}\quad or\quad \lambda (x,t)\lambda
_{x}(x,t)=\lambda _{t}(x,t)  \label{a1}
\end{equation}
Examples of particular solutions to this equation for particular
mathematical scenarions are exemplified as follows. \cite{witham} first
reported that the particular solution to the Monge nonlinear differential
equation (\ref{a1}) can be given as:

\begin{equation}
\lambda (x,t)=G(x+\lambda (x,t)t),\text{\cite{witham}}  \label{l}
\end{equation}
Leznov et al. \cite{leznov},{\bf \ }demonstrated later on that the free
motion Monge-type equations are intrinsically related to $X=F(\stackrel{.}{X,%
}X;t)$ when ODE $\stackrel{..}{X}=0$. Such function can be integrated
through quadratures, which allows the Monge differential equation's solution
to be presented as:

\begin{equation}
x-\lambda (x,t)t=f\langle \lambda (x,t)\rangle   \label{a2}
\end{equation}
Two further examples of trivial solutions to the Monge equation are as
follows:

\begin{equation}
x+t\lambda (x,t)=0\text{ and }\lambda (x,t)=0  \label{ap}
\end{equation}
Finally, Fairlie et. al found another solution to this equation in 1993 \cite
{fairlie}. This particular solution can be considered to be the extant
most-general solution. Nevertheless, it is not completely accurate, and thus
the need for more general, accurate forms to the Monge equation.

\begin{equation}
\lambda (x,t)=\frac{a+x}{1-t}  \label{l0}
\end{equation}
This solution can be reached upon using an arbitrary differentiable initial
value $F$ in terms of a Laplace or Fourier expansion as published in \cite
{fairlie}:

\begin{equation}
F(x)=c_{0}+c_{1}e^{\alpha x}+c_{2}e^{2\alpha x}+c_{3}e^{3\alpha x}+\cdots ,%
\text{ where }F(0)=c_{0}.  \label{l1}
\end{equation}
$\lambda (x,t)$ remains to be largely unknown despite of its wide number of
citations. We find a precise $\lambda (x,t)$ function through the thorough
use of 

basic differential geometry procedures. The order of a differential equation
is given by the maximum number of times the supposed unknown function has
been derived.

The Monge differential equation applications are exemplified in the Witham
book, in waves theory, and while obtaining the Bateman differential
equation. Fairlie et. al., solved the non linear differential equation of
the Monge Equation for special cases of their physical applications, mainly
in the field of hydrodynamic problems. He also generalized the equation into
a system of equations, which he called the Universal Field Equations \cite
{fairlie}, \cite{fairlie1}, \cite{fairlie2}, \cite{fairlie3}, \cite{fairlie4}%
. The dispersive deformations of the Monge Equation are studied using ideas
originating from topological quantum field theory and the deformation
quantization programme, \cite{strachan}, etc.$\ldots $.

\section{Statement of results}

We begin by standardizing notation and terminology. All results presented on
this paper are exact in the sense that integration constants depend only
upon the initial conditions. Such dependence will usually be pointed out.
The function $\lambda (x,t)$ is the unknown function and only depends upon $x
$ and $t$;

\begin{equation}
\lambda (x,t)=\sin \varphi (x,t)  \label{a3}
\end{equation}

Now substituting (\ref{a3}) into (\ref{a1}), the following equation  is
obtained;

\begin{equation}
\varphi _{x}(x,t)\sin \varphi (x,t)=\varphi _{t}(x,t)\text{ }or\text{ }\sin
\varphi (x,t)=\frac{\varphi _{t}(x,t)}{\varphi _{x}(x,t)}  \label{a4}
\end{equation}
Considering Figure 1,  (\ref{a4}) is derived with respect to $t$, to obtain
the following:

\begin{equation}
\varphi _{t}(x,t)\cos \varphi (x,t)=\frac{\partial \frac{\varphi _{t}(x,t)}{%
\varphi _{x}(x,t)}}{\partial t}=\left[ \frac{\varphi _{t}(x,t)}{\varphi
_{x}(x,t)}\right] _{t}  \label{a5}
\end{equation}
Consequently, we have:

\begin{equation}
\cos \varphi (x,t)=\frac{\varphi _{x}(x,t)\varphi _{tt}(x,t)-\varphi
_{tx}(x,t)\varphi _{t}(x,t)}{-\varphi _{t}(x,t)\varphi _{x}^{2}(x,t)}
\label{a6}
\end{equation}
See figure2.

Now, taking figure 2 into account and comparing it against Figure 1, the
following equations can be deduced:

\begin{equation}
\begin{array}{c}
\varphi _{x}(x,t)=-\varphi _{x}^{2}(x,t)\varphi _{t}(x,t)\text{ }%
\Longrightarrow  \\ 
\varphi _{x}(x,t)+_{x}^{2}(x,t)\varphi _{t}(x,t)=0\Longrightarrow  \\ 
\varphi _{x}(x,t)\left[ 1+\varphi _{x}(x,t)\varphi _{t}(x,t)\right] =0
\end{array}
\label{a8.a}
\end{equation}
Must be complete $\varphi _{x}(x,t)\neq 0$ then $\varphi _{x}(x,t)\varphi
_{t}(x,t)=-1$ because if $\varphi _{x}(x,t)=0$, then $\varphi (x,t)=f(t)$,
which will not satisfy the Monge equation (\ref{a1}),

\begin{equation}
\sqrt{\varphi _{x}^{2}(x,t)-\varphi _{t}^{2}(x,t)}=\varphi _{x}(x,t)\varphi
_{tt}(x,t)-\varphi _{tx}(x,t)\varphi _{t}(x,t)  \label{a8.b}
\end{equation}
and

\begin{equation}
\varphi _{t}(x,t)=\sqrt{\langle \varphi _{x}^{2}(x,t)\varphi
_{t}(x,t)\rangle ^{2}-\langle \varphi _{x}(x,t)\varphi _{tt}(x,t)-\varphi
_{tx}(x,t)\varphi _{t}(x,t)\rangle ^{2}}  \label{a8.c}
\end{equation}
Now substituting (\ref{a8.b}) into (\ref{a8.c}), and solving for $\varphi
_{t}(x,t)$:

\begin{equation}
\varphi _{x}^{4}(x,t)\varphi _{t}^{2}(x,t)=\varphi
_{x}^{2}(x,t)\Longrightarrow \ \varphi _{x}(x,t)\varphi _{t}(x,t)=\mp 1
\label{l2}
\end{equation}
One of the solutions does not exist:

\begin{equation}
\varphi _{x}(x,t)\varphi _{t}(x,t)=+1  \label{b1}
\end{equation}
Because $\varphi =\varphi (x,t)$ is a fuction dependant on $x$ and $t$,
anticonmutative properties\ with respect to the derivative of $x$ and $t$
can be deduced:

\begin{equation}
\frac{\partial \frac{\partial \varphi (x,t)}{\partial t}}{\partial x}=-\frac{%
\partial \frac{\partial \varphi (x,t)}{\partial x}}{\partial t}\text{ }or%
\text{ }\varphi _{xt}(x,t)=-\varphi _{tx}(x,t)  \label{l3}
\end{equation}
which can be proven, as shown in Appendix. If $\lambda (x,t)=\sin \varphi
(x,t)$, then

\begin{equation}
\lambda _{xt}(x,t)+\lambda _{tx}(x,t)=\left\{ \frac{\partial }{\partial x},%
\frac{\partial }{\partial t}\right\} \lambda (x,t)=\left\{ \frac{\partial }{%
\partial t},\frac{\partial }{\partial x}\right\} \lambda (x,t)=-2\lambda
(x,t)  \label{l4}
\end{equation}
The correct solution can then be presented as:

\begin{equation}
\varphi _{x}(x,t)\varphi _{t}(x,t)=-1  \label{l5}
\end{equation}
Which complies with equation (\ref{a8.a}) to deduce the following:

\begin{equation}
\varphi _{t}(x,t)=-\frac{1}{\varphi _{x}(x,t)}  \label{a9}
\end{equation}
When $\varphi _{xt}(x,t)=\varphi _{tx}(x,t)$, the commutative properties
remain unchanged. Now substituting (\ref{a9}) into (\ref{a4}):

\begin{equation}
\varphi _{t}(x,t)=i\sqrt{\sin \varphi (x,t)}\text{ and }\varphi _{x}(x,t)=%
\frac{-i}{\sqrt{\sin \varphi (x,t)}}  \label{a10}
\end{equation}
By deriving each (\ref{a10}) equation with respect to $x$ and $t$,
respectively, $\varphi _{xt}(x,t)=\frac{\partial \varphi _{t}(x,t)}{\partial
x}=i\frac{\cos \varphi (x,t)}{\sqrt{\sin \varphi (x,t)}}\varphi _{x}(x,t)$
and $\varphi _{tx}(x,t)=\frac{\partial \varphi _{x}(x,t)}{\partial t}=-i%
\frac{\cos \varphi (x,t)}{(\sin \varphi (x,t))^{\frac{3}{2}}}\varphi
_{t}(x,t)$ are obtained. If both equations are brought up together by
equaling them to one another, $\varphi _{xt}(x,t)=\varphi _{tx}(x,t)=i\frac{%
\cos \varphi (x,t)}{\sqrt{\sin \varphi (x,t)}}\varphi _{x}(x,t)=-i\frac{\cos
\varphi (x,t)}{(\sin \varphi (x,t))^{\frac{3}{2}}}\varphi _{t}(x,t)$, $%
\varphi _{x}(x,t)=\frac{\varphi _{t}(x,t)}{\sin \varphi (x,t)}$ is deduced,
which would pretty much be Eq.(\ref{a4}).,

Eq. (\ref{a10}) exactly represents two direct, first order differential
equations that are directly integrable:

\begin{equation}
\begin{array}{l}
\varphi (x,t)=i\int^{t}\sqrt{\sin \varphi (x,t^{\prime })}\text{ }dt^{\prime
}+g(x) \\ 
\text{ }\varphi (x,t)=-i\int^{x}\frac{dx^{\prime }}{\sqrt{\sin \varphi
(x^{\prime },t)}}+f(t)
\end{array}
\text{ }or  \label{pb3}
\end{equation}
Because both Eq. (\ref{pb3}) solutions are related ( i.e. through the
equation $\varphi _{x}(x,t)\varphi _{t}(x,t)=-1$) by the first derived
function with respect to $x$ and $t$, respectively, this means the solution
to the Monge equation must be unique. (\ref{pb3}) solution can be seen as
follows:

\begin{equation}
\begin{array}{l}
t+g(x)=\int^{\varphi (x,t)}\frac{d\varphi ^{\prime }(x,t)}{i\sqrt{\sin
\varphi ^{\prime }(x,t)}}=-2\sqrt{2}i{\bf F}\left\langle \alpha (x,t),\frac{1%
}{\sqrt{2}}\right\rangle \text{\quad where} \\ 
\alpha (x,t)=\arcsin \sqrt{\frac{2\sin \varphi (x,t)}{1+\sin \varphi
(x,t)+\cos \varphi (x,t)}}=\arcsin \sqrt{\frac{2\lambda (x,t)}{1+\lambda
(x,t)+\sqrt{1-\lambda ^{2}(x,t)}}}
\end{array}
\label{a12}
\end{equation}
or

\begin{equation}
\begin{array}{l}
x+f(t)=\int^{\varphi (x,t)}i\sqrt{\sin \varphi ^{\prime }(x,t)}d\varphi
^{\prime }(x,t)=-2\sqrt{2}i{\bf E}\left\langle \alpha _{1}(x,t),\frac{1}{%
\sqrt{2}}\right\rangle +\sqrt{2}i{\bf F}\left\langle \alpha _{1}(x,t),\frac{1%
}{\sqrt{2}}\right\rangle \quad \text{where} \\ 
\alpha _{1}(x,t)=\arcsin \left\langle \sqrt{2}\sin \frac{1}{2}(\frac{\pi }{2}%
-\varphi (x,t))\right\rangle =\arcsin \left\langle \sqrt{2}\sin \frac{1}{2}(%
\frac{\pi }{2}-\arcsin \lambda (x,t))\right\rangle 
\end{array}
\label{a14}
\end{equation}
see e.g. \cite{table}, where both solutions given by equations (\ref{a12})
and (\ref{a14}) are equivalent in $\lambda (x,t)$. $g(x)$ and $f(t)$ depend
only on the initial conditions, and cannot be zero or constant, because $%
\lambda (x,t)\lambda _{x}(x,t)=\lambda _{t}(x,t)$ must be satisfied.
Solutions (\ref{a12}) and (\ref{a14}) are implicit  to $\lambda (x,t)\lambda
_{x}(x,t)=\lambda _{t}(x,t)$. Finally, ${\bf F(\beta ,a)}$ and ${\bf E(\beta
,a)}$ are the Elliptic integrals of the first and second kind, respectively.

\subsection{\protect\appendix
 appendix}

In order to prove $\varphi _{xt}(x,t)=-\varphi _{tx}(x,t)$, using the
relation $\varphi _{x}(x,t)\varphi _{t}(x,t)=1$, $\varphi _{x}(x,t)=\varphi
_{x}^{2}(x,t)\varphi _{t}(x,t)$ must be fullfilled. Equation (\ref{a6}) must
be modified as it is indicated:

\begin{equation}
\cos \varphi (x,t)=\frac{\varphi _{x}(x,t)\varphi _{tt}(x,t)-\varphi
_{tx}(x,t)\varphi _{t}(x,t)}{-\varphi _{t}(x,t)\varphi _{x}^{2}(x,t)}=\frac{%
\varphi _{tx}(x,t)\varphi _{t}(x,t)-\varphi _{x}(x,t)\varphi _{tt}(x,t)}{%
\varphi _{t}(x,t)\varphi _{x}^{2}(x,t)}  \label{g1}
\end{equation}
see figure 3. Now taking into account Figure 3 and Figure1, we obtain:

\begin{equation}
\sqrt{\varphi _{x}^{2}(x,t)-\varphi _{t}^{2}(x,t)}=\varphi _{tx}(x,t)\varphi
_{t}(x,t)-\varphi _{x}(x,t)\varphi _{tt}(x,t)  \label{g2}
\end{equation}
from $\varphi _{x}(x,t)\varphi _{t}(x,t)=+1$ and Eq. (\ref{a4}), it is found
that:

\begin{equation}
\varphi _{x}(x,t)\varphi _{t}(x,t)=1\text{ then }\varphi _{t}(x,t)=\sqrt{%
\sin \varphi (x,t)}\text{and }\varphi _{x}(x,t)=\frac{1}{\sqrt{\sin \varphi
(x,t)}}  \label{g3}
\end{equation}
Now substituting (\ref{g3}) into (\ref{g2}):

\begin{equation}
\begin{array}{l}
\sqrt{\varphi _{x}^{2}(x,t)-\varphi _{t}^{2}(x,t)}=\sqrt{\frac{1}{\sin
\varphi (x,t)}-\sin \varphi (x,t)}=\sqrt{\frac{1-\sin ^{2}\varphi (x,t)}{%
\sin \varphi (x,t)}}= \\ 
=\sqrt{\frac{\cos ^{2}\varphi (x,t)}{\sin \varphi (x,t)}}=\frac{\cos \varphi
(x,t)}{\sqrt{\sin \varphi (x,t)}}\text{ now} \\ 
\varphi _{tx}(x,t)\varphi _{t}(x,t)-\varphi _{x}(x,t)\varphi _{tt}(x,t)=%
\frac{\cos \varphi (x,t)}{2\sqrt{\sin \varphi (x,t)}}-\frac{-\cos \varphi
(x,t)}{2\sin \varphi (x,t)}\sqrt{\sin \varphi (x,t)}= \\ 
=\frac{\cos \varphi (x,t)}{\sqrt{\sin \varphi (x,t)}}\text{ is complete (\ref
{g2}), where} \\ 
\varphi _{tt}(x,t)=\frac{\cos \varphi (x,t)}{2\sqrt{\sin \varphi (x,t)}}%
\varphi _{t}(x,t)=\frac{\cos \varphi (x,t)}{2}\text{ and} \\ 
\text{ }\varphi _{tx}(x,t)=\frac{\partial \varphi _{x}(x,t)}{\partial t}=%
\frac{\partial \frac{1}{\sqrt{\sin \varphi (x,t)}}}{\partial t}= \\ 
-\frac{\cos \varphi (x,t)}{2\langle \sin \varphi (x,t)\rangle ^{\frac{3}{2}}}%
\varphi _{t}(x,t)=-\frac{\cos \varphi (x,t)}{2\sin \varphi (x,t)}
\end{array}
\label{g4}
\end{equation}

From figure 3 and figure 1, we found:

\begin{equation}
\varphi _{t}(x,t)=\sqrt{\langle \varphi _{x}^{2}(x,t)\varphi
_{t}(x,t)\rangle ^{2}-\langle \varphi _{x}(x,t)\varphi _{tt}(x,t)-\varphi
_{tx}(x,t)\varphi _{t}(x,t)\rangle ^{2}}  \label{g6}
\end{equation}
Now substituting (\ref{g3}) into (\ref{g6}) and using the results of (\ref
{g4}):

\begin{equation}
\begin{array}{l}
\varphi _{t}(x,t)=\sqrt{\sin \varphi (x,t)}\text{ }=\sqrt{\langle \varphi
_{x}^{2}(x,t)\varphi _{t}(x,t)\rangle ^{2}-\langle \varphi _{x}(x,t)\varphi
_{tt}(x,t)-\varphi _{xt}(x,t)\varphi _{t}(x,t)\rangle ^{2}}\text{ } \\ 
=\sqrt{\frac{\sin \varphi (x,t)}{\sin ^{2}\varphi (x,t)}-\frac{\cos
^{2}\varphi (x,t)}{\sin \varphi (x,t)}}=\sqrt{\frac{1-\cos ^{2}\varphi (x,t)%
}{\sin \varphi (x,t)}}=\sqrt{\sin \varphi (x,t)}\text{ is complete}
\end{array}
\label{l6}
\end{equation}

Now we found $\varphi _{xt}(x,t)$ and proved that $\varphi
_{xt}(x,t)=-\varphi _{tx}(x,t)$ is as follows:

\begin{equation}
\begin{array}{l}
\varphi _{xt}(x,t)=\text{ }\frac{\partial \varphi _{t}(x,t)}{\partial x}=%
\text{ }\frac{\partial \sqrt{\sin \varphi (x,t)}}{\partial x}=\frac{\cos
\varphi (x,t)}{2\sqrt{\sin \varphi (x,t)}}\varphi _{x}(x,t)= \\ 
=\frac{\cos \varphi (x,t)}{2\sqrt{\sin \varphi (x,t)}}\frac{1}{\sqrt{\sin
\varphi (x,t)}}=\frac{\cos \varphi (x,t)}{2\sin \varphi (x,t)}\text{ but} \\ 
\varphi _{tx}(x,t)=-\frac{\cos \varphi (x,t)}{2\sin \varphi (x,t)}\text{
containen in (\ref{g4}) then} \\ 
\varphi _{xt}(x,t)=-\varphi _{tx}(x,t)\text{ is proved}
\end{array}
\label{g5}
\end{equation}

Now we prove that the second mixed partial derivative of $\lambda (x,t)$ is
not independent from the order of differentiation for this case (i.e. that $%
\lambda _{xt}(x,t)+\lambda _{tx}(x,t)=-2\lambda (x,t)$). From (\ref{a3}), we
have

\begin{equation}
\begin{array}{l}
\lambda (x,t)=\sin \varphi (x,t)\text{ then }\lambda _{x}(x,t)=\varphi
_{x}\cos \varphi (x,t)\text{ and }\lambda _{t}(x,t)=\varphi _{t}\cos \varphi
(x,t)\text{ next} \\ 
\lambda _{xt}(x,t)=-\varphi _{x}\varphi _{t}\sin \varphi (x,t)+\varphi
_{xt}\cos \varphi (x,t)\text{ and }\lambda _{tx}(x,t)=-\varphi _{t}\varphi
_{x}\sin \varphi (x,t)+\varphi _{tx}\cos \varphi (x,t) \\ 
\text{ taking the sum of }\lambda _{xt}(x,t)\text{ and }\lambda _{tx}(x,t)%
\text{ it is obtained} \\ 
\lambda _{xt}(x,t)+\lambda _{tx}(x,t)=-\varphi _{x}\varphi _{t}\sin \varphi
(x,t)-\varphi _{t}\varphi _{x}\sin \varphi (x,t)\text{ } \\ 
\text{where we used }\varphi _{tx}=-\text{ }\varphi _{xt}\text{ and we know
that }\varphi _{x}\varphi _{t}=\varphi _{x}\varphi _{t}=1\text{ }\text{then}
\\ 
\lambda _{xt}(x,t)+\lambda _{tx}(x,t)=-2\sin \varphi (x,t)\text{ now we
using (\ref{a3}), finally }\lambda _{xt}(x,t)+\lambda _{tx}(x,t)=-2\lambda
(x,t)
\end{array}
\label{l7}
\end{equation}
This is the reason why equation (\ref{b1}), is not a solution to the Monge
Equation (\ref{a1}).

\subsection{figures}

\begin{figure}[htbp]
	\centering
		\includegraphics[width=0.50\textwidth]{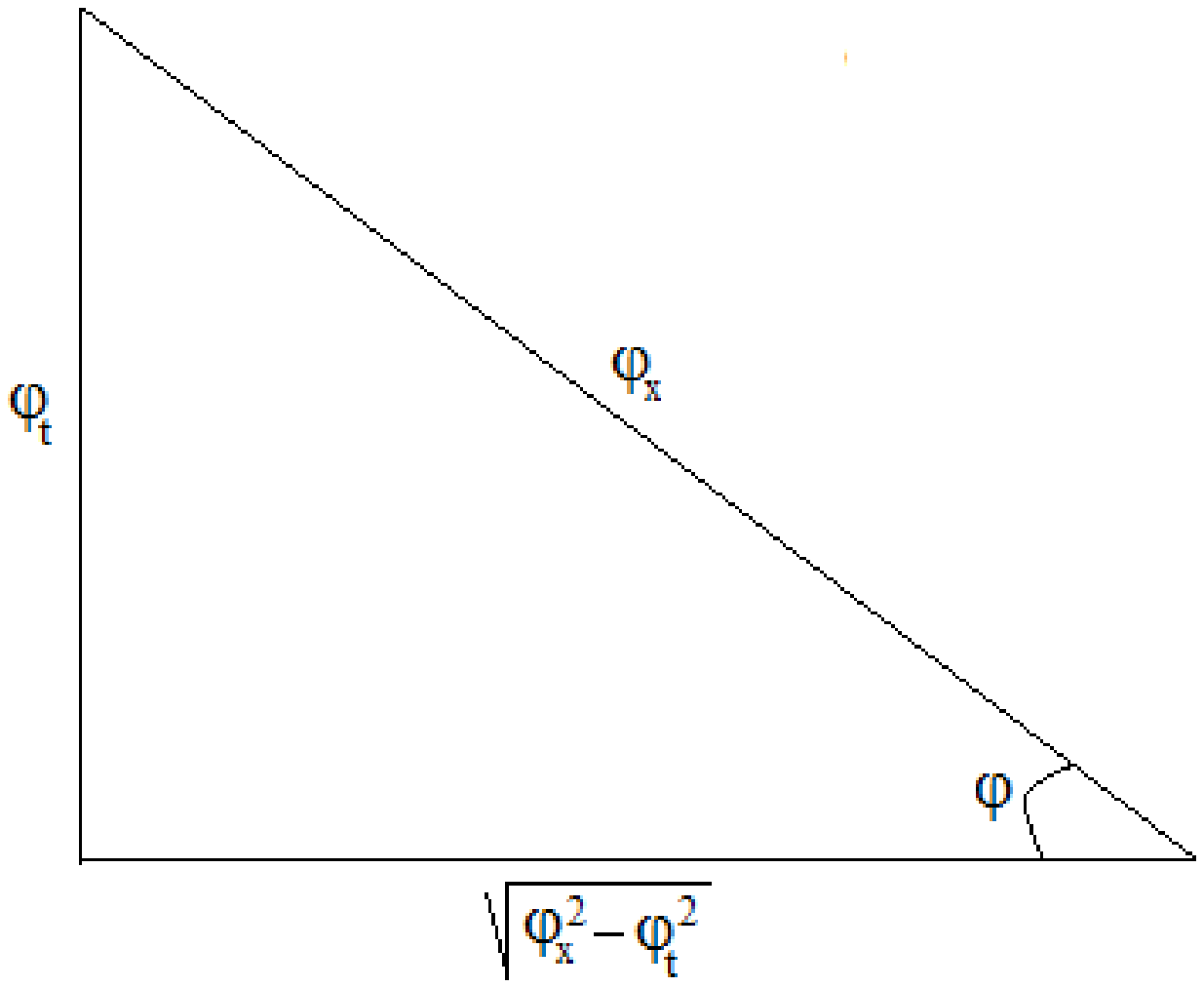}
	\caption{Figure 1}
	\label{fig:figure1}
\end{figure}

\begin{figure}[htbp]
	\centering
		\includegraphics[width=0.60\textwidth]{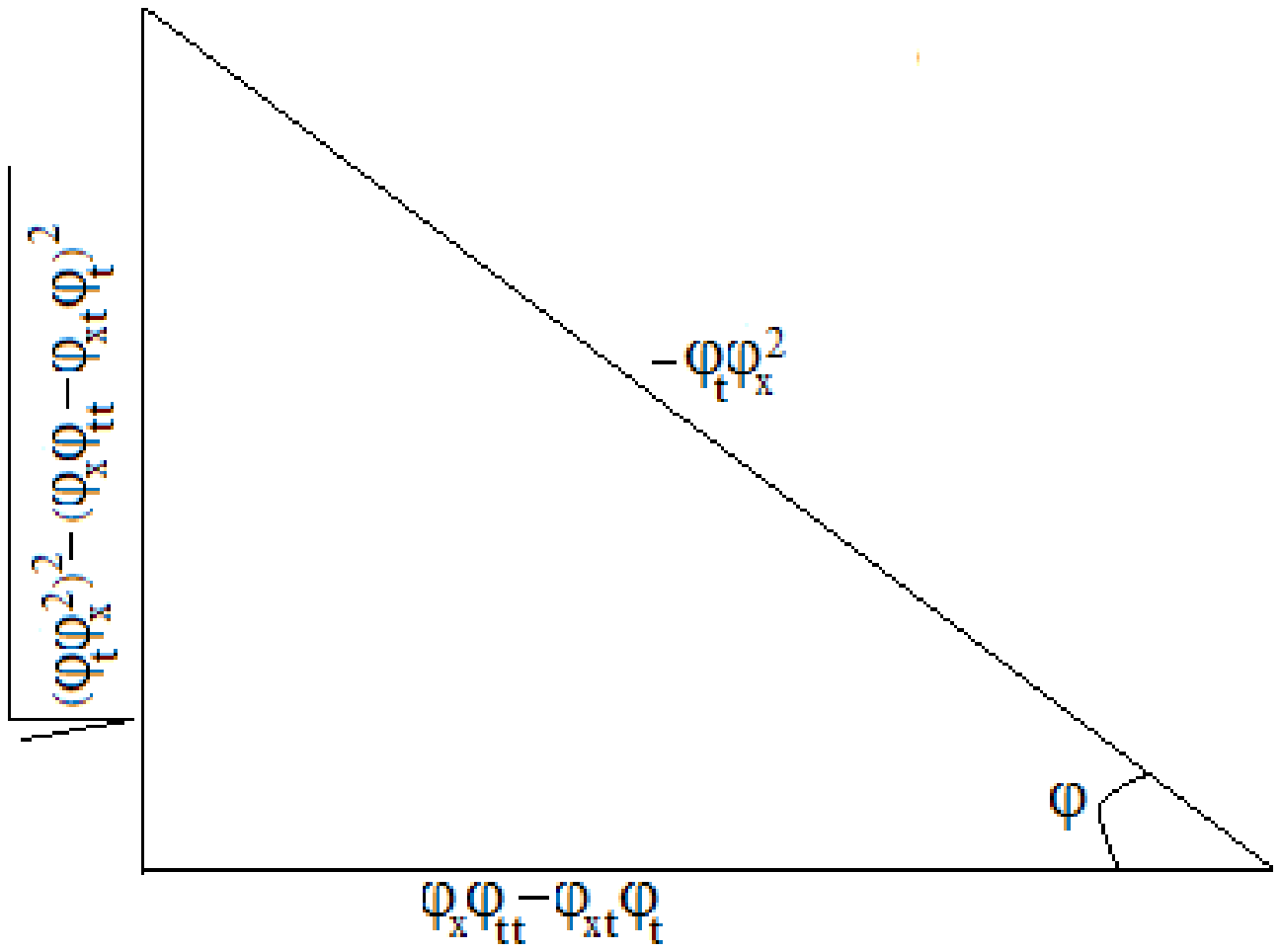}
	\caption{Figure 2}
	\label{fig:figure2}
\end{figure}

\begin{figure}[htbp]
	\centering
		\includegraphics[width=0.70\textwidth]{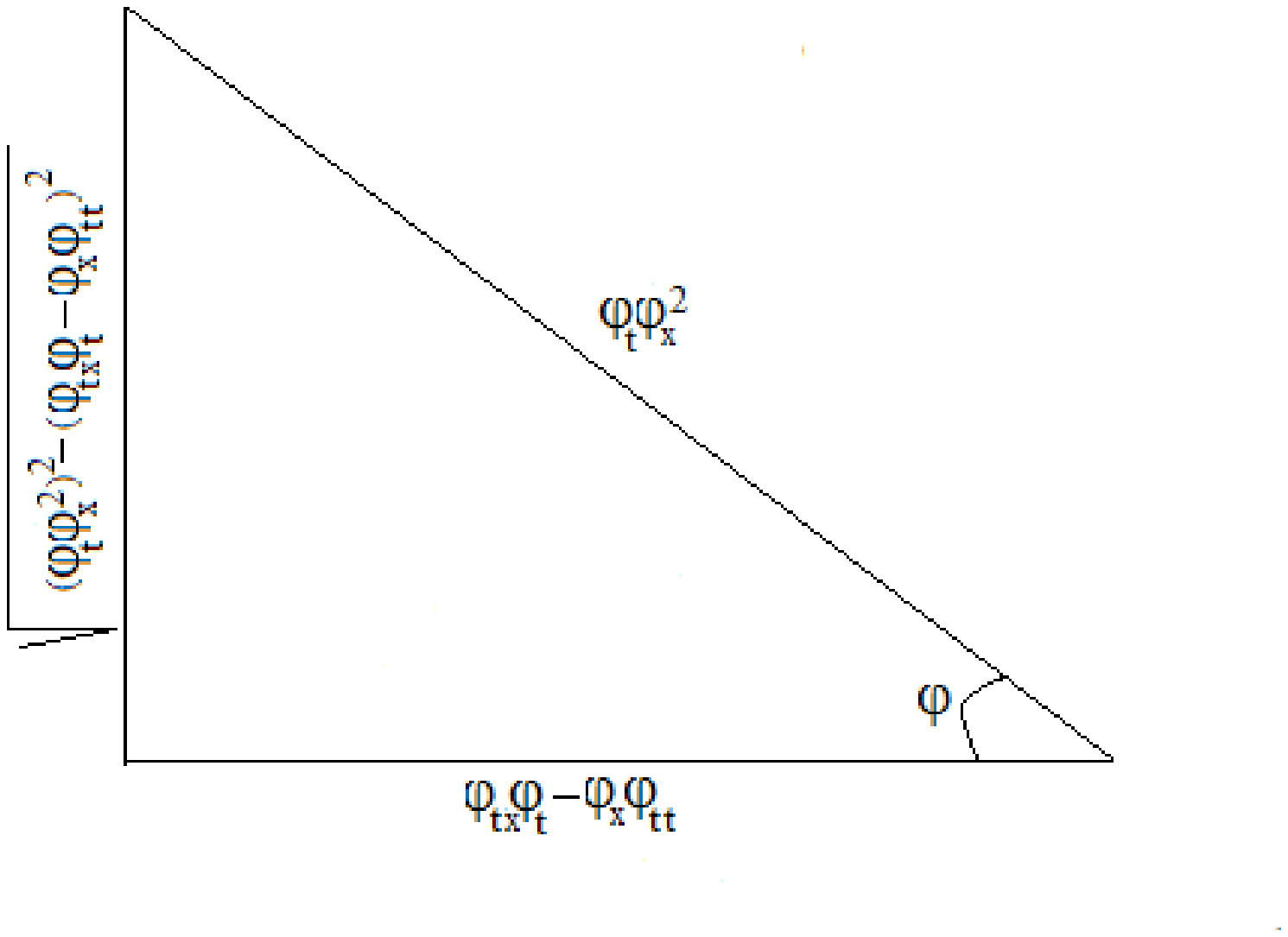}
	\caption{Figure 3}
	\label{fig:figure3}
\end{figure}

\end{document}